# Self-Bootstrapping Automated Program Repair: Using LLMs to Generate and Evaluate Synthetic Training Data for Bug Repair

**David de-Fitero-Domínguez[1], Antonio García-Cabot[1,*], Eva García-López[1]**

[1]*Departamento de Ciencias de la Computación, Universidad de Alcalá, Alcalá de Henares, Madrid, Spain*

*Corresponding author: a.garciac@uah.es*

## 1 Abstract

This paper presents a novel methodology for enhancing Automated Program Repair (APR) through synthetic data generation utilizing Large Language Models (LLMs). Current APR systems are constrained by the limited availability of high-quality training data encompassing diverse bug types across multiple programming languages. The proposed approach addresses this limitation through a two-phase process: a synthetic sample generation followed by a rigorous quality assessment. Multiple state-of-the-art LLMs were employed to generate approximately 30,000 paired examples of buggy and fixed code across 12 programming languages and 13 bug categories. Subsequently, these samples underwent cross-model evaluation against five criteria: correctness, code quality, security, performance, and completeness. Experimental evaluation on the VulRepair test set dataset showed statistically significant improvements in Perfect Prediction rates, with the quality-filtered synthetic dataset achieving 17.18% (Top@1) and 23.00% (Top@5) compared to the baseline's 11.68% and 18.88% respectively, representing a 47% relative improvement in Top@1 and 22% in Top@5. The methodology was validated through rigorous statistical testing, including ANOVA and post-hoc Tukey's Honest Significant Difference analysis. Furthermore, the best-performing configurations surpassed existing systems despite using a less computationally intensive decoding strategy. This research establishes a self-bootstrapping paradigm in which LLMs generate and evaluate their own training data, suggesting promising directions for addressing data scarcity in similar software engineering tasks and advancing the development of robust, adaptable tools for automated code maintenance.

# 2 Introduction

Software maintenance is a major expense in the software development process, costing billions of dollars every year (Jorgensen & Shepperd, 2007; Weiss et al., 2007). A big part of this cost comes from the challenge of finding and fixing software issues, including both general bugs that affect functionality and security vulnerabilities that can put systems at risk (Erlingsson et al., 2010; Gao et al., 2018). In languages like C and C++, which are often used for their performance benefits, these issues can be particularly difficult to manage. Developers have to deal with everything from logic errors to memory corruption problems like buffer overflows and use-after-free bugs. Fixing these issues manually is a slow, error-prone, and exhausting process, which is why automated approaches are becoming so important (Britton et al., 2020; Erlingsson et al., 2010).

APR is designed to make fixing software issues easier by automatically generating patches for bugs and vulnerabilities (Zhang et al., 2024). Traditional methods, such as those based on heuristics (Jiang et al., 2018; Martinez & Monperrus, 2016; Yuan & Banzhaf, 2020), templates (Koyuncu et al., 2020; Liu et al., 2019), or constraints (Durieux & Monperrus, 2016; Martinez & Monperrus, 2018; Xuan et al., 2017) have been useful, but they often face limitations when dealing with complex or diverse bug scenarios. These approaches are constrained by their reliance on predefined patterns or rules, which can limit their effectiveness in addressing the wide variety of bugs encountered in real-world software development (Xia et al., 2023). The rise of deep learning, and especially LLMs, offers a fresh approach. These models, trained on large amounts of code data, are better at understanding the context of bugs and generating fixes that are more accurate and reliable (Islam et al., 2024; Zhou, Cao, et al., 2024).

The success of LLM-based APR systems depends heavily on the quality of their training data. Unfortunately, many existing datasets are too small, lack variety, or do not provide enough information about the context of the bugs and fixes (Li & Guo, 2024; Prenner & Robbes, 2023). This makes it hard for models to learn how to deal with a wide range of bugs and vulnerabilities, limiting their usefulness in real-world scenarios. To unlock the full potential of LLMs for program repair, we need better, larger, and more diverse datasets that can cover the full spectrum of software issues.

Therefore, the limitations of existing datasets highlight the importance of exploring synthetic data generation. The ability to use LLMs to generate diverse, realistic, and contextually rich training examples, as explored in recent works (Kim et al., 2024; Liu et al., 2024; Taori et al., 2023) (unrelated to APR), offers a promising path to overcome these data limitations. By using LLMs to create synthetic datasets specifically tailored for program repair tasks, we can potentially overcome the issues of relying just on existing, limited datasets and unlock the full potential of LLMs to address a wider range of software bugs and vulnerabilities. This approach not only provides a scalable alternative to manual data collection but also allows us to develop more versatile and robust APR systems.

In this work, we introduce a method for creating high-quality synthetic training data using LLMs. Our approach leverages state-of-the-art LLMs to generate pairs of buggy and fixed code, and then evaluates these pairs with the same models to filter out low-quality samples. This methodology ensures the resulting dataset remains realistic, diverse, and focused on producing useful training data. The generation of high-quality synthetic examples, rather than relying on existing limited datasets, effectively addresses data scarcity problems while improving overall data quality.

To validate our method, we applied it to the preprocessed dataset from Vulrepair (Fu et al., 2022), which combines CVEFixes (Bhandari et al., 2021) and BigVul (Fan et al., 2020), and demonstrated that our approach improves the performance of other LLM-based APR systems. Our experimental results show that quality-filtered synthetic data improves repair accuracy beyond what is achievable through training with manually curated datasets. Specifically, our best models outperform existing state-of-the-art systems like VulMaster (Zhou, Kim, et al., 2024) and VulRepair (Fu et al., 2022).

The main contributions of this work are: (1) A novel methodology for generating and evaluating synthetic data for APR using multiple LLMs, with cross-model evaluation to ensure quality; (2) Empirical evidence demonstrating that quality-filtered synthetic data outperforms both unfiltered data and manually collected training samples; and (3) A rigorous statistical validation framework that confirms the significance of our results across multiple experimental configurations, establishing more robust evaluation standards for APR research.

# 3 Related work

APR techniques are commonly divided into four main categories: search-based, constraint-based, template-based, and learning-based approaches (Xia et al., 2023; Zhang et al., 2023, 2024). Each paradigm has evolved to address specific challenges in automated software repair, with varying degrees of success across different bug types and programming contexts.

Search-based techniques, such as GenProg (Le Goues et al., 2012) and more recent SimFix (Jiang et al., 2018), focus on exploring a predefined patch space using heuristics, code mutations, and similarity-based searches. These methods are effective in identifying and fixing various types of bugs but often face challenges with large search spaces, computational inefficiency, and the risk of generating incorrect or irrelevant patches (Huang et al., 2023). They also rely heavily on test cases, which may not always be available or comprehensive enough to guide effective repairs (Zhang et al., 2023).

Constraint-based approaches, including tools like SemFix (Nguyen et al., 2013), DirectFix (Mechtaev et al., 2015), and Angelix (Mechtaev et al., 2016), convert the repair problem into a constraint-solving task. These methods utilize formal specifications or test cases to guide the repair process, effectively reducing search space and improving patch quality. However, their dependency on accurate and specific constraints limits their flexibility, making them suitable primarily for simpler or well-defined defects (Monperrus, 2018).

Template-based approaches, such as PAR (Kim et al., 2013), FixMiner (Koyuncu et al., 2020), and TBar (Liu et al., 2019), use predefined patterns or templates to generate repairs. These methods are efficient and can produce high-quality patches for common defect types, but they struggle with rare or complex bugs that fall outside their predefined templates (Martinez & Monperrus, 2016). Despite their limitations, template-based approaches remain valuable for addressing recurring bug patterns.

Learning-based APR has made significant progress by leveraging Deep Learning (DL) techniques to automatically learn bug-fixing patterns from large codebases. Early efforts, such as the work of Tufano et al. (2019), framed APR as a Neural Machine Translation (NMT) task, treating buggy code as input to be "translated" into its correct form. While Recurrent neural network (RNNs) models (Rumelhart et al., 1986), such as SequenceR (Z. Chen et al., 2021), highlighted the potential of data-driven approaches, they faced limitations in capturing long-range code dependencies and ensuring syntactic correctness. Subsequent approaches, like DLFix (Li et al., 2020), addressed these challenges by incorporating Abstract Syntax Trees (ASTs) to improve the ability of the model to represent code structure.

The introduction of LLMs has transformed APR by overcoming many of the limitations of earlier approaches (Zhang et al., 2024). Unlike RNN-based methods, LLMs built on the Transformer architecture (Vaswani et al., 2017) effectively capture long-range dependencies and incorporate contextual understanding of diverse programming languages (Feng et al., 2020; Lu et al., 2021; Wang et al., 2021). Early examples include TFix (Berabi et al., 2021), which fine-tuned LLMs for text-to-text repair, and AlphaRepair (Xia & Zhang, 2022), which used cloze-style approaches for generating patches by filling masked code segments.

LLMs demonstrate exceptional versatility through various learning paradigms. Fine-tuning allows the models to adjust their parameters to improve performance on specific repair tasks. This technique has been widely adopted in works such as those from Berabi et al. (2021), Chen et al. (2023), Fu et al. (2022) and W. Wang et al. (2023), demonstrating its effectiveness in handling diverse repair challenges, from fixing security vulnerabilities to addressing static warnings.

Beyond fine-tuning, few-shot learning enables LLMs to generalize from a minimal number of examples directly provided in the input prompts, requiring no additional training. Studies like that of Nashid et al. (2023) showcase the utility of few-shot learning in APR tasks, particularly in scenarios where data availability is limited. Zero-shot learning further extends this adaptability by requiring no task-specific training or examples. Notable work includes ChatRepair (Xia & Zhang, 2023, 2024), which employs conversational feedback loops to iteratively refine patches, leveraging failure information to guide repairs, and ContrastRepair (Kong et al., 2024), which complements this process by integrating positive feedback from passing tests, improving the context provided to the model for debugging.

The adaptability of LLMs has allowed APR to expand into underexplored bug types and programming languages, addressing gaps left by traditional repair techniques. For example, VulRepair (Fu et al., 2022) fine-tunes CodeT5 (Wang et al., 2021) to address software vulnerabilities, while RTLFixer (Tsai et al., 2024) employs iterative prompting to fix Verilog hardware bugs, extending APR capabilities beyond traditional software languages. InferFix (Jin et al., 2023), targeting static warnings, integrates semantic retrieval and augmented prompts to tackle issues detected by static analyzers. These works illustrate the versatility of LLMs in APR, enabling repairs in scenarios where traditional techniques are ineffective, such as rare languages (e.g., Rust (Deligiannis et al., 2023)) or complex domains like Web UI repair (Xu et al., 2024).

Despite these advances, the effectiveness of LLM-based APR systems remains contingent on the quality and diversity of available training data. Many existing datasets are too small, lack variety, or fail to provide sufficient contextual information about bugs and their fixes (Li & Guo, 2024; Prenner & Robbes, 2023). Recent work has begun exploring synthetic data generation as a potential solution, with approaches like those of Chaudhary (2023), S. Kim et al. (2024), R. Liu et al. (2024), and Taori et al. (2023), demonstrating the potential of using LLMs to create training examples for various tasks. However, these approaches typically focus on general instruction-following capabilities rather than domain-specific challenges like APR.

Within APR specifically, several approaches have explored synthetic or curated data generation. SelfAPR (Ye et al., 2023) generates synthetic bugs by applying predefined perturbation rules to correct code, validates them through compilation and test execution, and trains models to reverse these perturbations. While effective, this approach is project-specific and limited to Java. InstructRepair (Fu et al., 2025) takes a different approach by instruction-tuning models on existing bug-fix datasets, enriching them with structured bug information integrated into instruction templates. However, it relies on the availability of existing datasets and does not include a quality filtering mechanism.

Our work builds upon these foundations and introduces a comprehensive methodology for generating and evaluating synthetic data specifically tailored for APR tasks. Rather than relying on predefined perturbation rules or existing bug-fix datasets, our approach prompts LLMs to generate entirely new bug-fix pairs across diverse bug categories and 12 programming languages, without requiring project-specific compilation or test infrastructure. The generated samples are then assessed through a multi-criteria, cross-model quality filtering mechanism using six diverse LLMs. This approach directly addresses the fundamental data limitations that have constrained previous methods, enhancing the capabilities of LLM-based APR systems across diverse programming languages and bug types. Through careful quality filtering and evaluation, we demonstrate that synthetic data can effectively supplement or even replace traditional training datasets, opening new possibilities for more versatile and effective automated program repair.

# 4 Methodology

For the generation and filtering of synthetic data, we use LLMs in a dual role. First as generators of buggy/fixed code pairs, then as evaluators of those same examples. This creates a cycle where models both create and assess training data, establishing a novel paradigm for synthetic data generation.

The process begins with synthetic data generation, where we prompt different LLMs to generate examples covering 12 programming languages and 13 bug categories. Each example includes a description of the bug, the buggy code, and its fixed version. Subsequently, the same LLMs evaluate each generated example using multiple criteria: correctness, code quality, security, performance, and completeness. We apply a quality threshold of 8.5 to retain only high-quality examples.

After filtering, we preprocess the data by formatting the examples consistently for model training, using a specific representation of bugs and fixes. We then fine-tune the Qwen 2.5 Coder 7B model (Hui et al., 2024; Yang et al., 2024) using different combinations of datasets to measure the impact of our synthetic data. Finally, we test all configurations on the VulRepair test set (Fu et al., 2022) and employ rigorous statistical analysis to determine whether differences in performance are statistically significant. Our goal is to demonstrate that high-quality synthetic data can supplement or even replace traditional datasets, providing a scalable solution to data limitations in the field.

## 4.1 Synthetic code generation

Our synthetic code generation methodology leverages state-of-the-art LLMs to create high quality training data. The framework integrates multiple frontier models, including NVIDIA's *Llama-3.1-Nemotron-70B-Instruct-HF* (Wang et al., 2024), Meta's *Meta-Llama-3.1-70B-Instruct* (Dubey et al., 2024), *Qwen2.5-72B-Instruct* (Hui et al., 2024; Yang et al., 2024), *Mistral-Small-Instruct-2409*[1], *Gemma-2-27B-it* (Team et al., 2024), and *Qwen2.5-32B-Instruct* (Hui et al., 2024; Yang et al., 2024). While these models may be characterized as general-purpose, they incorporate a substantial amount of code in their pretraining data to improve logical reasoning capabilities, and demonstrate strong performance on code generation benchmarks such as HumanEval (M. Chen et al., 2021) and MBPP (Austin et al., 2021), with strong support for multilingual programming (Jiang et al., 2025). Additionally, using models from different organizations with variations in training data and fine-tuning approaches is essential for our cross-model evaluation strategy, as this diversity helps reduce potential biases that could arise from relying on models from a single organization.

For efficient inference, we use the vLLM engine (Kwon et al., 2023), which implements paged attention and efficient memory management. We set the temperature to 0.7 to balance diversity with controlled randomness (Tunstall et al., 2022). Combined with different random seeds for each model run, these settings promote varied yet coherent code examples. Also, the architectural differences between models and their distinct training datasets further contribute to varied outputs, minimizing the likelihood of generating identical or highly similar bug-fix pairs across different LLMs. The generation pipeline randomly samples from 12 programming languages (Python, Java, C++, Go, Ruby, Rust, Swift, Kotlin, PHP, C#, JavaScript, and Pascal) and 13 bug types (including off-by-one errors, infinite loops, resource leaks, concurrency issues, improper error handling, memory management errors, and security vulnerabilities). Each model generates 5,000 samples, creating a representative dataset across multiple programming contexts.

The generation process follows three structured steps. First, we construct prompts that instruct models to generate bug examples with their fixes, incorporating randomly selected bug types and programming languages. The required output format includes error descriptions, buggy code, and corrected solutions structured with XML-style tags (<error_description>, <buggy_code>, and <fixed_code>). Second, we submit these prompts to LLMs with instructions to emulate expert software developers, encouraging realistic bug generation and effective fixes while maintaining structural consistency. Third, we validate outputs through automated pattern matching, filtering out malformed responses to ensure only properly structured examples proceed.

This approach yields approximately 30,000 synthetic samples with balanced distribution across programming languages (8.09-8.61% each, Table 1) and bug categories (7.50-7.90% each, Table 2), ensuring comprehensive coverage without bias. High validity rates achieved by most models (several reaching 100% valid outputs, Table 3) demonstrate the robustness of our prompt design and LLM capabilities.

[1] https://huggingface.co/mistralai/Mistral-Small-Instruct-2409

| Language | Count | Percentage |
|---|---|---|
| C# | 2504 | 8.35% |
| C++ | 2514 | 8.38% |
| Go | 2428 | 8.09% |
| Java | 2582 | 8.61% |
| JavaScript | 2466 | 8.22% |
| Kotlin | 2488 | 8.29% |
| PHP | 2567 | 8.56% |
| Pascal | 2499 | 8.33% |
| Python | 2510 | 8.37% |
| Ruby | 2445 | 8.15% |
| Rust | 2443 | 8.14% |
| Swift | 2554 | 8.51% |

*Table 1: Initial samples per programming language*

| Bug | Count | Percentage |
|---|---|---|
| Arithmetic errors (e.g., division by zero or integer overflows) | 2267 | 7.56% |
| Concurrency issues (e.g., race conditions, deadlocks) | 2361 | 7.87% |
| Improper error handling (e.g., missing exceptions or uncaught exceptions) | 2370 | 7.90% |
| Inconsistent state in object-oriented code (e.g., failing to update object attributes correctly) | 2290 | 7.63% |
| Incorrect conditionals (e.g., wrong boolean expressions) | 2361 | 7.87% |
| Incorrect loop boundaries | 2282 | 7.61% |
| Incorrect use of API methods or functions | 2249 | 7.50% |
| Infinite loops | 2358 | 7.86% |
| Memory leaks or buffer overflows (for C/C++ code) | 2310 | 7.70% |
| Misuse of data structures (e.g., incorrect initialization or access) | 2264 | 7.55% |
| Off-by-one errors | 2357 | 7.86% |
| Resource leaks (e.g., open files or sockets not being closed) | 2274 | 7.58% |
| SQL injections or unsafe user input handling | 2257 | 7.52% |

*Table 2: Initial samples per bug type*

| Model | Valid samples | Validity (%) |
|---|---|---|
| Llama-3.1-Nemotron-70B-Instruct-HF | 4829 | 96.58% |
| Meta-Llama-3.1-70B-Instruct | 5000 | 100.00% |
| Mistral-Small-Instruct-2409 | 4990 | 99.80% |
| Qwen2.5-32B-Instruct | 5000 | 100.00% |
| Qwen2.5-72B-Instruct | 5000 | 100.00% |
| gemma-2-27b-it | 4998 | 99.96% |

*Table 3: Initial samples per model*

## 4.2 Scoring and ranking

Quality evaluation of synthetic samples is fundamental to our methodology, ensuring that only high-quality samples proceed to the training phase. We evaluate each generated sample across multiple dimensions (correctness, code quality, security, performance, and completeness), on a 0-10 scale. Our cross-model evaluation approach uses the same frontier LLMs used for sample generation as evaluators, with each model evaluating the entire corpus of approximately 30,000 samples. This mutual evaluation strategy mitigates individual model biases and provides a more objective quality assessment framework than could be achieved by single model evaluation.

The evaluation process compares the buggy code with its fixed version to generate a “diff” (similar to what version control systems produce) showing changed, added, or removed lines. This diff representation is included in a carefully designed evaluation prompt that asks the model to act as an expert software developer and evaluate the patch against the predetermined criteria. We use a temperature setting of 0.2 during evaluation (versus 0.7 for generation) to promote more consistent scoring. Analysis of 1,000 samples across temperatures 0.2, 0.5, and 0.7 (five repetitions each, three evaluator models) confirmed this choice: temperatures 0.5 and 0.7 exhibited 128.2% and 205.0% higher within-model variance, respectively, compared to temperature 0.2 (avg std = 0.0883 vs 0.2015 and 0.2693, respectively), with this pattern consistent across all evaluators (Llama-3.1-Nemotron-70B-Instruct-HF, Mistral-Small-Instruct-2409, and Gemma-2-27b-it). Additionally, we use the Outlines library (Willard & Louf, 2023) to enforce a structured JSON output format with multiple fields for each evaluation criterion.

It is important to note that our methodology establishes a self-bootstrapping paradigm where the same LLMs that generate synthetic samples also serve as evaluators. Specifically, each of the six models evaluates the entire dataset, regardless of which model originally generated each sample. This cross-model evaluation means that each sample receives six independent quality assessments.

The evaluation relies entirely on the understanding of code quality of the LLMs, that is, we do not employ additional syntactic validators, compilers, or external tools to verify the assessments. Instead, we trust the capabilities of the models to provide informed judgments about bug realism and fix correctness and then use consensus across multiple evaluators to ensure reliability. The strong performance improvements observed in our experiments (Section 5) validate this design choice, demonstrating that LLM-based evaluation effectively identifies high-quality training samples.

As additional validation, we performed syntax validation on a stratified sample of 2,500 synthetic examples across five programming languages (Python, JavaScript, Ruby, PHP, and Go) using language-specific parsers. This post-hoc analysis showed that 96.6% of unfiltered samples were syntactically valid, improving to 97.7% after applying the quality threshold (≥8.5), with per-language validity ranging from 96.0% (JavaScript) to 99.8% (PHP) on filtered samples. The small proportion of invalid samples (2.3%) primarily consisted of code snippets lacking full program context such as import statements. This validation confirms that LLM-based quality filtering effectively retains syntactically correct code without requiring explicit syntactic validators in the evaluation pipeline.

To assess the reliability of LLM-based evaluation, we conducted a small human validation study on a stratified sample of 100 synthetic examples. Samples were stratified by inter-model agreement (variance in scores across the six evaluators) to ensure representation across quality levels: high-agreement samples with both high and low scores, medium-agreement samples, and low-agreement samples with high variance. An expert evaluator assessed these samples using the same five quality criteria on a 0-10 scale, with weighted scores calculated identically to the LLM evaluation. The results show strong correlation between LLM consensus scores and human expert assessment, with Pearson correlations averaging r=0.666 across all criteria. The weighted overall scores showed even stronger correlation (r=0.737). Notably, samples stratified by agreement level exhibited clear discrimination in human quality assessment: high-agreement high-score samples received mean human scores of 8-9, while low-agreement samples received substantially lower scores in the 2-4 range (Table 4). This suggests that consensus-based filtering effectively identifies high-quality training data.

The validation also revealed that LLM evaluators assign systematically higher scores compared to human assessment (mean LLM score 7.68 vs human 5.54, difference of 2.14 points), using a narrower range of the scoring scale. However, despite this optimistic bias, LLM scores preserve the relative quality ordering: high-quality samples according to human judgment also score high with LLMs, and low-quality samples score correspondingly lower. Most importantly, at our chosen threshold of 8.5, LLM-based filtering retained 22% of samples while human evaluation would retain 20%, with both approaches agreeing on the retain/discard decision for 86% of examples. This supports both the threshold selection and suggests that LLM-based quality assessment, despite systematic score inflation, remains effective for practical data curation when combined with cross-model consensus.

| Group | Human score | LLMs score |
|---|---|---|
| **High agreement, high score (n=25)** | 8.8 ± 1.3 | 9.1 ± 0.4 |
| **High agreement, low score (n=25)** | 4.7 ± 2.0 | 7.7 ± 0.4 |
| **Medium agreement (n=25)** | 5.6 ± 2.4 | 8.4 ± 0.3 |
| **Low agreement (n=25)** | 2.6 ± 2.7 | 6.1 ± 1.4 |

*Table 4: Human validation results. Mean weighted scores and standard deviation for each stratified group.*

Finally, individual scores are combined into a single quality score using weighted averages. Our weighting scheme prioritizes correctness (0.3) as the fundamental requirement - a fix must actually resolve the bug. Code quality and length (0.2 each) ensure solutions are maintainable and substantive, with length scores using logarithmic transformation to reward comprehensive examples without bias toward excessive verbosity. Security, performance, and completeness (0.1 each) capture additional quality dimensions. This distribution reflects practical priorities in code repair: functional correctness is essential, but good engineering practices and comprehensive solutions also matter for real-world applicability.

This multi-dimensional weighted scoring approach aligns with established practices in automated code assessment (Parvathy et al., 2025), where program correctness is typically weighted highest, followed by structural aspects and other quality indicators. We then apply a quality threshold of 8.5 to determine which samples to include in our final dataset. This threshold was chosen as a practical compromise (strict enough to filter out lower quality examples, while still retaining approximately 70% of the original samples for training).

#### *4.2.1.1 Overall aggregation and pre-filtering analysis*

After the generation phase, our dataset contained 29,646 unique synthetic samples. Analysis of the quality scores from all the evaluator models showed an average score of 8.58 with a standard deviation of 0.86.

Table 5 shows the average scores received by samples from each generator model. The score patterns reveal interesting characteristics of our consensus-based filtering approach. For instance, samples generated by *Llama-3.1-Nemotron-70B-Instruct-HF* achieved the highest mean scores (8.84) with lowest variability (std dev 0.29), while *gemma-2-27b-it* received the lowest average score (8.16) and showed the most variability (std dev 1.30). When acting as evaluators, models showed inverse patterns: *Mistral-Small-Instruct-2409* assigned the highest average score (9.27), while *Llama-3.1-Nemotron-70B-Instruct-HF* gave the lowest (8.09), suggesting that it applied stricter evaluation standards.

These patterns reflect different evaluation stringency levels that contribute to our consensus-based filtering. The variation in both generation and evaluation patterns demonstrates the value of using multiple models, as it ensures diverse perspectives in both creating and filtering samples, preventing any single model's biases from dominating the dataset.

| Evaluated model | Mean score | Std dev |
|---|---|---|
| Llama-3.1-Nemotron-70B-Instruct-HF | **8.84** | **0.29** |
| Qwen2.5-72B-Instruct | 8.80 | 0.62 |
| Meta-Llama-3.1-70B-Instruct | 8.78 | 0.42 |
| Qwen2.5-32B-Instruct | 8.67 | 0.66 |
| Mistral-Small-Instruct-2409 | 8.24 | 1.12 |
| gemma-2-27b-it | 8.16 | 1.30 |

*Table 5: Average scores by evaluated model (before filtering). The highest mean is bolded, as is the lower std deviation.*

Programming language analysis (Table 6) shows C# achieving the highest average quality score (8.81), followed by C++ (8.69), while Python and Rust samples received the lowest scores (8.43 and 8.37, respectively). These differences may reflect varying exposure to high-quality code examples in certain languages during pre-training, or inherent language characteristics affecting bug fix clarity. Among bug types (Table 7), SQL injection vulnerabilities scored the highest (8.95), likely due to well-established fix patterns, while arithmetic errors (8.35) and especially concurrency issues (8.23) scored lowest, reflecting their inherent complexity.

| Language | Mean score | Std dev |
|---|---|---|
| C# | **8.81** | **0.64** |
| C++ | 8.69 | 0.66 |
| Go | 8.67 | 0.81 |
| Java | 8.66 | 0.73 |
| JavaScript | 8.52 | 0.88 |
| Kotlin | 8.55 | 0.77 |
| PHP | 8.62 | 0.66 |

| Language | Mean score | Std dev |
|---|---|---|
| Pascal | 8.68 | 0.77 |
| Python | 8.43 | 1.00 |
| Ruby | 8.44 | 0.97 |
| Rust | 8.37 | 1.23 |
| Swift | 8.53 | 0.94 |

*Table 6: Average scores by programming language. Highest mean and lowest std dev are shown in bold*

| Bug Type | Mean score | Std dev |
|---|---|---|
| Arithmetic errors (e.g., division by zero or integer overflows) | 8.35 | 0.57 |
| Concurrency issues (e.g., race conditions, deadlocks) | 8.23 | 0.97 |
| Improper error handling (e.g., missing exceptions or uncaught exceptions) | 8.58 | 0.32 |
| Inconsistent state in object-oriented code (e.g., failing to update object attributes correctly) | 8.47 | 0.93 |
| Incorrect conditionals (e.g., wrong boolean expressions) | 8.61 | 1.01 |
| Incorrect loop boundaries | 8.86 | 1.04 |
| Incorrect use of API methods or functions | 8.49 | 0.87 |
| Infinite loops | 8.56 | 0.94 |
| Memory leaks or buffer overflows | 8.49 | 0.63 |
| Misuse of data structures (e.g., incorrect initialization or access) | 8.69 | 0.51 |
| Off-by-one errors | 8.76 | 1.19 |
| Resource leaks (e.g., open files or sockets not being closed) | 8.51 | 0.99 |
| SQL injections or unsafe user input handling | **8.95** | **0.40** |

*Table 7: Average scores by bug. Highest mean and lowest std dev are shown in bold*

#### *4.2.1.2 Post-filtering analysis*

After applying our quality threshold of 8.5, the dataset was reduced to 20,832 samples (70% retention). This threshold effectively removed outliers with unusually low scores while preserving the majority of high-quality examples, as evidenced by the increased average quality score from 8.58 to 8.94 and a reduced standard deviation from 0.86 to 0.26.

Model contributions to the filtered dataset (Table 8) aligned with how well they performed during the pre-filtering evaluation phase. *Llama-3.1-Nemotron-70B-Instruct-HF* contributed the most samples (21.07%), followed by *Qwen2.5-72B-Instruct* (18.93%) and *Meta-Llama-3.1-70B-Instruct* (18.75%). Smaller models contributed fewer high-quality samples: *Mistral-Small-Instruct-2409* (12.15%) and *gemma-2-27b-it* (12.05%), consistent with their lower pre-filtering scores (Table 5).

| Evaluated Model | Contribution to the final dataset |
|---|---|
| Llama-3.1-Nemotron-70B-Instruct-HF | **21.07%** |
| Qwen2.5-72B-Instruct | 18.93% |
| Meta-Llama-3.1-70B-Instruct | 18.75% |

| Evaluated Model | Contribution to the final dataset |
|---|---|
| Qwen2.5-32B-Instruct | 17.05% |
| Mistral-Small-Instruct-2409 | 12.15% |
| gemma-2-27b-it | 12.05% |

*Table 8: Percentage of samples provided by each model to the final dataset. Highest contribution is shown in bold*

Post-filtering analysis reveals more uniform quality across programming languages (Table 9), with scores ranging from 8.89 (Python and Ruby) to 9.01 (C#). Some languages remained particularly strong in the filtered dataset, with C# (9.56%), Pascal (9.03%), and C++ (8.79%) having the highest proportions of samples passing the quality threshold. This suggests that the proposed models were particularly effective at generating high quality examples in these languages.

| Language | Mean score | Count | Percentage |
|---|---|---|---|
| C# | **9.01** | **1992** | **9.56%** |
| C++ | 8.93 | 1832 | 8.79% |
| Go | 8.99 | 1794 | 8.61% |
| Java | 8.97 | 1810 | 8.69% |
| JavaScript | 8.91 | 1627 | 7.81% |
| Kotlin | 8.92 | 1571 | 7.54% |
| PHP | 8.91 | 1761 | 8.45% |
| Pascal | 8.96 | 1881 | 9.03% |
| Python | 8.89 | 1621 | 7.78% |
| Ruby | 8.89 | 1555 | 7.46% |
| Rust | 8.92 | 1624 | 7.80% |
| Swift | 8.93 | 1764 | 8.47% |

*Table 9: Final average scores by programming language (filtered). Highest values are shown in bold*

Bug type analysis (Table 10), shows that well-defined problems achieved highest scores: *Incorrect loop boundaries* and *Off-by-one errors* both scored 9.13, reflecting their straightforward fix patterns. Complex issues such as *Arithmetic Errors* (8.72) and *Concurrency Issues* (8.77) scored lower. Interestingly, fixes for *SQL injections or unsafe user input handling* not only scored well (9.02) but also made up the largest portion of the filtered dataset (10.20%), suggesting that the chosen models are particularly effective at generating high-quality fixes for security vulnerabilities.

| Bug Type | Mean Score | Count | Percentage |
|---|---|---|---|
| Arithmetic errors (e.g., division by zero or integer overflows) | 8.72 | 997 | 4.79% |
| Concurrency issues (e.g., race conditions, deadlocks) | 8.77 | 1187 | 5.70% |
| Improper error handling (e.g., missing exceptions or uncaught exceptions) | 8.75 | 1524 | 7.32% |
| Inconsistent state in object-oriented code (e.g., failing to update object attributes correctly) | 8.89 | 1545 | 7.42% |
| Incorrect conditionals (e.g., wrong boolean expressions) | 9.01 | 1756 | 8.43% |

| Bug Type | Mean Score | Count | Percentage |
|---|---|---|---|
| Incorrect loop boundaries | **9.13** | 2009 | 9.64% |
| Incorrect use of API methods or functions | 8.84 | 1498 | 7.19% |
| Infinite loops | 8.96 | 1705 | 8.18% |
| Memory leaks or buffer overflows | 8.85 | 1279 | 6.14% |
| Misuse of data structures (e.g., incorrect initialization or access) | 8.88 | 1696 | 8.14% |
| Off-by-one errors | **9.13** | 1996 | 9.58% |
| Resource leaks (e.g., open files or sockets not being closed) | 8.99 | 1515 | 7.27% |
| SQL injections or unsafe user input handling | 9.02 | **2125** | **10.20%** |

*Table 10: Final average scores by bug type (filtered). The highest values are in bold.*

## 4.3 Data preprocessing

The training and evaluation data for our experiments originate from VulRepair (Fu et al., 2022), which preprocessed and combined samples from the CVEFixes (Bhandari et al., 2021) and BigVul (Fan et al., 2020) datasets. We use the published train/test split from VulRepair to ensure comparability with prior work. We carefully curated the datasets to eliminate duplicate entries and removed overlapping samples between training and test sets to prevent overfitting. The test dataset remained unchanged to ensure fair comparison.

Building on our previous work (de-Fitero-Dominguez et al., 2024), we improved the data format with improved special tokens usage for clearer delineation of error descriptions, vulnerable lines, and source code.

The input format structures prompts with specific tokens: **<inst>** marks the instruction block start; **<desc>** introduces the error description; **<file>** specifies the filename; **<lines>** lists vulnerable line numbers; numbered source code follows; and **</inst>** ends the block.

Below is a template representing the input format:

```
<inst>
<desc>{Error description here}
<file>{File name}
<lines>{line1 line2 ... lineN}
{1    buggy code line 1
2    buggy code line 2
...  ...}
</inst>
```

The output format generates patches as diffs rather than complete corrected code, improving efficiency and precision. It uses: **<file>** for filename; **<le>** to mark line ranges; modified code lines only (unchanged lines are not included); and **<sep>** to separate multiple change blocks.

Below is a template representing the output format:

```
<file>{FileName}
{start_line}<le>{end_line}
```

```
{modified code line 1
modified code line 2
...}
<sep>
... (additional hunks if any)
```

The intuition behind this format is to provide both a comprehensive and structured problem representation. The input format provides a clear bug description, precise vulnerability locations, and complete source code context, allowing the model to understand the issues without unnecessary repetition. Meanwhile, the output format focuses solely on changes through a diff that clearly marks affected line ranges and separates multiple blocks of changes. Together, these formats enable the model to generate accurate, targeted code patches.

## 4.4 Model fine-tuning

For our fine-tuning experiments, we selected the Qwen 2.5 Coder 7B model (Hui et al., 2024; Yang et al., 2024), a state-of-the-art 7-billion-parameter model known for its outstanding code understanding and generation capabilities. Our experimental configurations systematically evaluate synthetic data impact:

- (Baseline – *vulrep*)*:* VulRepair training set only
- (1 - *vulrep_synt_85*): Baseline + filtered synthetic dataset (threshold 8.5)
- (2 - *vulrep_commitpack*): Baseline + CommitPackFT
- (3 - *vulrep_synt_85_commitpack*): Baseline + filtered synthetic + CommitPackFT
- (4 - *vulrep_synt_full*): Baseline + unfiltered synthetic dataset

Our baseline configuration uses only the VulRepair training data (Fu et al., 2022), with no test set overlap. Note that the VulRepair training data consists predominantly of C/C++ code vulnerabilities, so the reported metrics specifically reflect the effectiveness of the models in repairing the C/C++ vulnerabilities despite our multi-language synthetic data generation.

The integration of the CommitPackFT dataset (Muennighoff et al., 2023) was a critical component of our experimental setup. This 2GB filtered subset of the larger 4TB CommitPack collection contains commits from 350 programming languages and has been carefully curated to include only high-quality commit messages that resemble natural language instructions. Its proven effectiveness in improving APR performance (de-Fitero-Dominguez et al., 2024) makes it an ideal benchmark for comparing against our synthetic data approach.

For all experiments, we used LoRA (Low-Rank Adaptation) (Hu et al., 2021) for efficient fine-tuning with the following hyperparameters: three epochs, learning rate of 3e-4 using a cosine learning rate scheduler, batch size of 256, and maximum sequence length of 4,096 tokens. LoRA configuration followed Dettmers et al. (2023) recommendations: rank (r) of 16, LoRA alpha of 32, and dropout rate of 0.1, applied to all linear projection layers. Internal testing confirmed these settings provided stable performance with minimal sensitivity to parameter variations.

## 4.5 Model evaluation

We evaluate our approach using the Perfect Prediction (PP%) metric, which considers a generated patch correct if it exactly matches the ground truth repair. We evaluate performance over different training configurations under two evaluation settings: Top@1, where we generate a single patch per test instance (temperature 0.4 for more deterministic output), and Top@5, where we generate five patches and determine success by the presence of a correct match in one of them (temperature of 0.8 for diversity). Due to the inherent stochasticity of the sampling process, we repeat each experiment 50 times to ensure that the results remain stable and are not affected by chance. We perform all evaluations on the test set from VulRepair (Fu et al., 2022), enabling direct comparison with previous work including VulMaster (Zhou, Kim, et al., 2024), VulRepair and our previous work using Mistral (de-Fitero-Dominguez et al., 2024). Our primary goal is not to outperform existing systems, but rather to investigate the feasibility of synthetic data generation and to validate our methodology by cross-evaluating LLMs.

To determine statistical significance of performance differences across configurations, we use rigorous statistical testing. We use an analysis of variance (ANOVA) test (Fisher, 1992), to compare mean performance across multiple experimental conditions simultaneously. Prior to Anova, we verify required assumptions using the Shapiro-Wilk test for normality (SHAPIRO & WILK, 1965) and Levene's test for homogeneity of variances (Levene & others, 1960), both suitable for our moderate sample size and distribution characteristics.

Following ANOVA confirmation of significant differences, we perform Tukey's Honest Significant Difference (HSD) test for pairwise comparisons. This post-hoc analysis identifies specific configurations yielding meaningful performance improvements while controlling family-wise error rate (Tukey, 1949). This comprehensive statistical framework provides conclusive evidence of the effectiveness of synthetic data generation in improving APR systems, and validates the viability of our quality filtering methodology, establishing a more rigorous empirical foundation for future research in this area.

# 5 Experimental results

In this section, we present the results of our experimental evaluation, following the methodology outlined above. Our experiments are designed to address three guiding research questions:

- RQ1: Does quality-filtered synthetic data improve APR performance compared to training on real-world data alone?
- RQ2: Is quality filtering necessary, or does more data inherently lead to better results?
- RQ3: How sensitive is the approach to filtering parameters (threshold, criteria weights)?

We analyze the performance of each training configuration using the PP% metric under both the Top@1 and Top@5 settings. Section 5.1 reports the PP% results addressing

RQ1 and RQ2, Section 5.2 validates these differences through statistical testing, and Section 5.3 presents the ablation study addressing RQ3.

## 5.1 Performance results

This section addresses RQ1 and RQ2 by comparing the performance of models trained with filtered synthetic data, unfiltered synthetic data, and the original baseline. Table 11 summarizes the Perfect Prediction rates achieved by each training configuration under both Top@1 and Top@5 evaluation settings. Each experiment was repeated 50 times to ensure reliability, and we report the mean performance along with standard deviations.

As shown in Table 11, all configurations using synthetic data outperformed the baseline (*vulrep*) in both Top@1 and Top@5 settings. Notably, the quality-filtered synthetic dataset (*vulrep_synt_85*) achieved substantially better results than the unfiltered version (*vulrep_synt_full*), with an improvement of approximately 2 percentage points in both evaluation settings. This supports our hypothesis that data quality is more important than quantity for effective APR. The combination of filtered synthetic data with CommitPackFT (*vulrep_synt_85_commitpack*) yielded the highest performance in the Top@1 setting, while the filtered synthetic data alone (*vulrep_synt_85*) performed best in the Top@5 scenario.

| Configuration | Top@1 Mean PP% | Top@1 Std dev | Top@5 Mean PP% | Top@5 Std dev |
|---|---|---|---|---|
| *vulrep (baseline)* | 11.68% | 0.0039 | 18.88% | 0.0047 |
| *vulrep_commitpack* | 16.14% | **0.0033** | 21.79% | **0.0038** |
| *vulrep_synt_85* | 17.18% | 0.0036 | **23.00%** | 0.0043 |
| *vulrep_synt_85_commitpack* | **17.26%** | 0.0040 | 22.47% | 0.0045 |
| *vulrep_synt_full* | 15.21% | 0.0042 | 21.13% | 0.0048 |

*Table 11: Mean PP% and Std dev over 50 runs on the VulRepair test set. Highest mean and lowest std dev are marked in bold*

With respect to RQ1, these results suggest that quality-filtered synthetic data can improve APR performance over training on real-world data alone: *vulrep_synt_85* achieves a 47% relative improvement in Top@1 (17.18% vs 11.68%) and a 22% relative improvement in Top@5 (23.00% vs 18.88%) compared to the baseline. Regarding RQ2, the approximately 2 percentage points difference between the filtered and unfiltered versions in both evaluation settings suggests that quality filtering plays an important role, and that data quality may matter more than quantity for effective APR training.

To validate whether these observed differences are statistically significant, we conducted rigorous statistical testing, which is explained in detail in the following section. First, we verified the basic assumptions required for analysis of variance (ANOVA), then performed the analysis followed by post-hoc pairwise comparisons.

## 5.2 Statistical validation

To confirm the significance of the differences observed for RQ1 and RQ2, we conducted comprehensive statistical analysis using rigorous testing procedures. This analysis validates that the observed performance improvements are statistically significant and not due to random variation.

### 5.2.1 Assumption testing: normality and homogeneity of variances

Before applying ANOVA, we rigorously tested the basic statistical assumptions using the Shapiro-Wilk test for normality and Levene's test for homogeneity of variances. Table 12 shows the results of the normality tests, while Table 13 presents the results of the homogeneity of variances test.

| Configuration | p-value (top@1) | p-value (top@5) |
|---|---|---|
| *vulrep* | 0.3818 | 0.7078 |
| *vulrep_commitpack* | 0.8015 | 0.6665 |
| *vulrep_synt_85* | 0.1352 | 0.9496 |
| *vulrep_synt_85_commitpack* | 0.3236 | 0.0932 |
| *vulrep_synt_full* | 0.2200 | 0.8093 |

*Table 12: Shapiro-Wilk test for checking normality (Top@1 and Top@5)*

All configurations successfully passed the normality test ($p > 0.05$ in all cases), confirming that our data distributions do not deviate significantly from normality.

We also performed Levene's test to verify the homogeneity of variances across configurations, with results presented in Table 13. With both key assumptions (normality and homogeneity of variances) successfully met, we proceeded with ANOVA to assess significant differences between training configurations.

| Evaluation setting | p-value |
|---|---|
| Top@1 | 0.3060 |
| Top@5 | 0.7086 |

*Table 13: Levene's test for homogeneity of variances*

### 5.2.2 Analysis of variance (ANOVA)

Our ANOVA results (Table 14) revealed compelling evidence of significant differences between the evaluated configurations in both Top@1 ($F=1735.92$, $p<0.0001$) and Top@5 ($F=646.27$, $p<0.0001$) settings. These extremely low p-values ($p<0.0001$) provide strong statistical evidence that there are genuine differences in perfect prediction performance among the tested configurations. Given these conclusive findings, we conducted a post-hoc analysis using Tukey's HSD test to precisely identify which configurations yielded statistically significant improvements.

| Evaluation setting | F-Statistic | p-value |
|---|---|---|
| Top@1 | 1735.92 | <0.0001 |
| Top@5 | 646.27 | <0.0001 |

*Table 14: ANOVA results*

### 5.2.3 Post-hoc analysis: Tukey's HSD test

After confirming statistically significant differences by ANOVA, we performed a comprehensive post-hoc analysis using Tukey's HSD test. This approach allows for pairwise comparisons while controlling for family-wise error rate (FWER), effectively minimizing false positives that could result from multiple comparisons. Table 15 and Table 16 present the detailed results for the Top@1 and Top@5 evaluation settings, respectively.

| Group 1 | Group 2 | Mean diff. | p-value |
|---|---|---|---|
| *vulrep* | *vulrep_commitpack* | 0.0446 | **0.000** |
| *vulrep* | *vulrep_synt_85* | 0.0549 | **0.000** |
| *vulrep* | *vulrep_synt_85_commitpack* | **0.0558** | **0.000** |
| *vulrep* | *vulrep_synt_full* | 0.0352 | **0.000** |
| *vulrep_commitpack* | *vulrep_synt_85* | 0.0104 | **0.000** |
| *vulrep_commitpack* | *vulrep_synt_85_commitpack* | 0.0112 | **0.000** |
| *vulrep_commitpack* | *vulrep_synt_full* | -0.0093 | **0.000** |
| *vulrep_synt_85* | *vulrep_synt_85_commitpack* | 0.0009 | 0.806 |
| *vulrep_synt_85* | *vulrep_synt_full* | -0.0197 | **0.000** |
| *vulrep_synt_85_commitpack* | *vulrep_synt_full* | -0.0206 | **0.000** |

*Table 15: Tukey's HSD test results (top@1). Highest mean difference and values with p < 0.05 are shown in bold*

| Group 1 | Group 2 | Mean diff. | p-value |
|---|---|---|---|
| *vulrep* | *vulrep_commitpack* | 0.0291 | **0.000** |
| *vulrep* | *vulrep_synt_85* | **0.0411** | **0.000** |
| *vulrep* | *vulrep_synt_85_commitpack* | 0.0359 | **0.000** |
| *vulrep* | *vulrep_synt_full* | 0.0225 | **0.000** |
| *vulrep_commitpack* | *vulrep_synt_85* | 0.0121 | **0.000** |
| *vulrep_commitpack* | *vulrep_synt_85_commitpack* | 0.0069 | **0.000** |
| *vulrep_commitpack* | *vulrep_synt_full* | -0.0066 | **0.000** |
| *vulrep_synt_85* | *vulrep_synt_85_commitpack* | -0.0052 | **0.000** |
| *vulrep_synt_85* | *vulrep_synt_full* | -0.0187 | **0.000** |
| *vulrep_synt_85_commitpack* | *vulrep_synt_full* | -0.0134 | **0.000** |

*Table 16: Tukey's HSD test results (top@5). Highest mean difference and values with p < 0.05 are shown in bold*

To better interpret these statistical results, we visualized probability density distributions for each experimental configuration. Figure 1 and Figure 2 illustrate the distributions of PP% scores across all training configurations. These visualizations provide strong confirmation of our statistical analysis (configurations with statistically significant differences identified by Tukey's test show clear separation in their distributions). Conversely, configurations without significant differences, such as *vulrep_synt_85* vs. *vulrep_synt_85_commitpack* in Top@1-show overlapping distributions, reinforcing the consistency between visual patterns and statistical results.

These results provide statistical confirmation for the answers to RQ1 and RQ2 presented in Section 5.1, establishing that the observed improvements are not attributable to random variation.

In Figure 2, we also include reference performance benchmarks from previous research: VulMaster (20.0%) (Zhou, Kim, et al., 2024), VulRepair (16.8%) (Fu et al., 2022), and our previous work using Mistral (22.04%) (de-Fitero-Dominguez et al., 2024). It is important to clarify that these are external performance benchmarks evaluated on the VulRepair test set, and are distinct from our *vulrep* baseline configuration, which was trained using the VulRepair training data using the same model as the other configurations. These benchmarks are included only in the Top@5 evaluation, as they used beam search with 50 candidates, making Top@5 a more appropriate (though still conservative) comparison point than Top@1. We specifically use the VulRepair results as reported in the VulMaster study rather than from the original VulRepair paper, as VulMaster identified and corrected a data leakage issue where the original VulRepair evaluation dataset contained samples that overlapped with its training data, artificially inflating performance metrics. The corrected evaluation by VulMaster provides a more reliable benchmark for comparison.

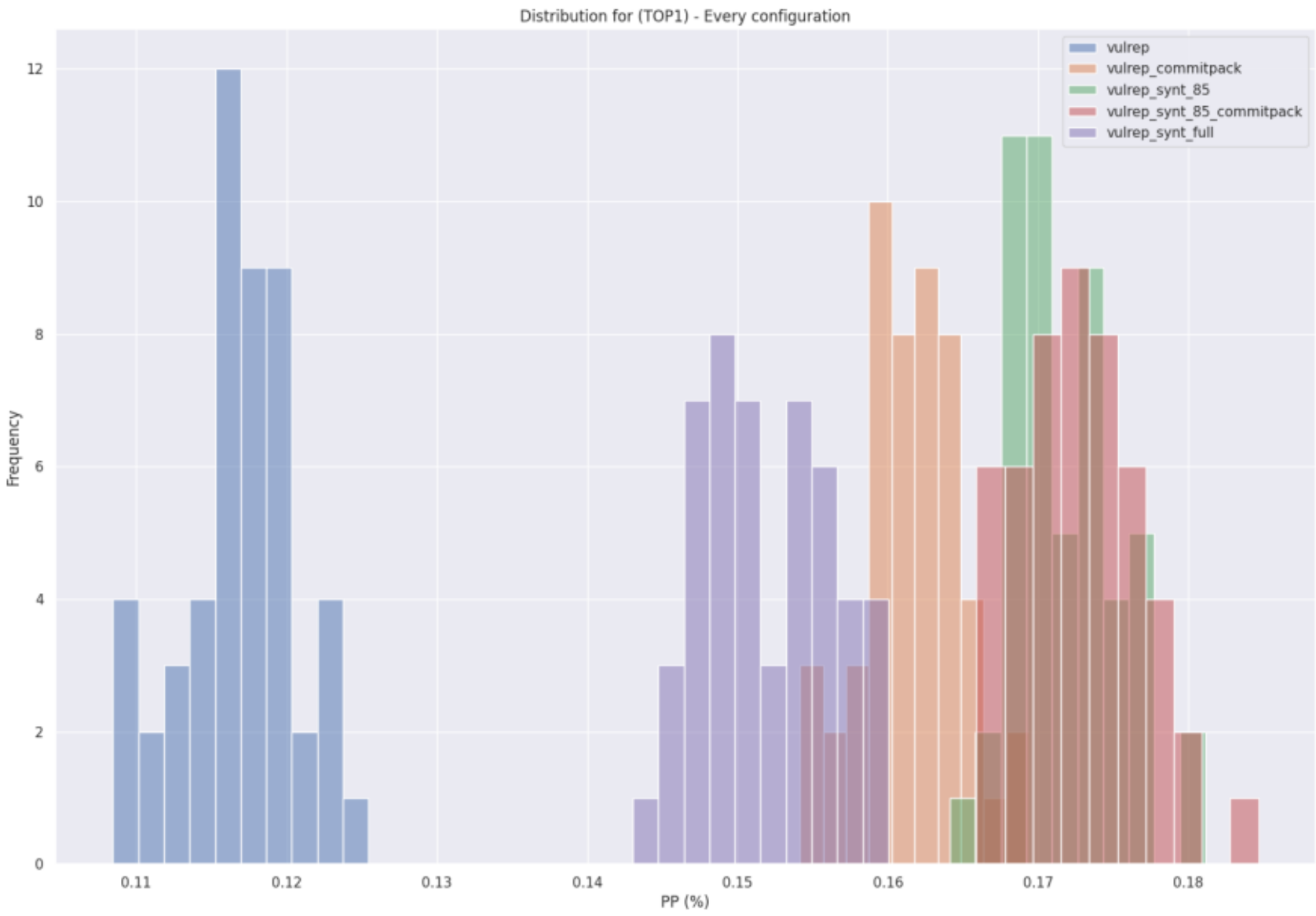

*Figure 1: Probability density distributions for Top@1 configurations*

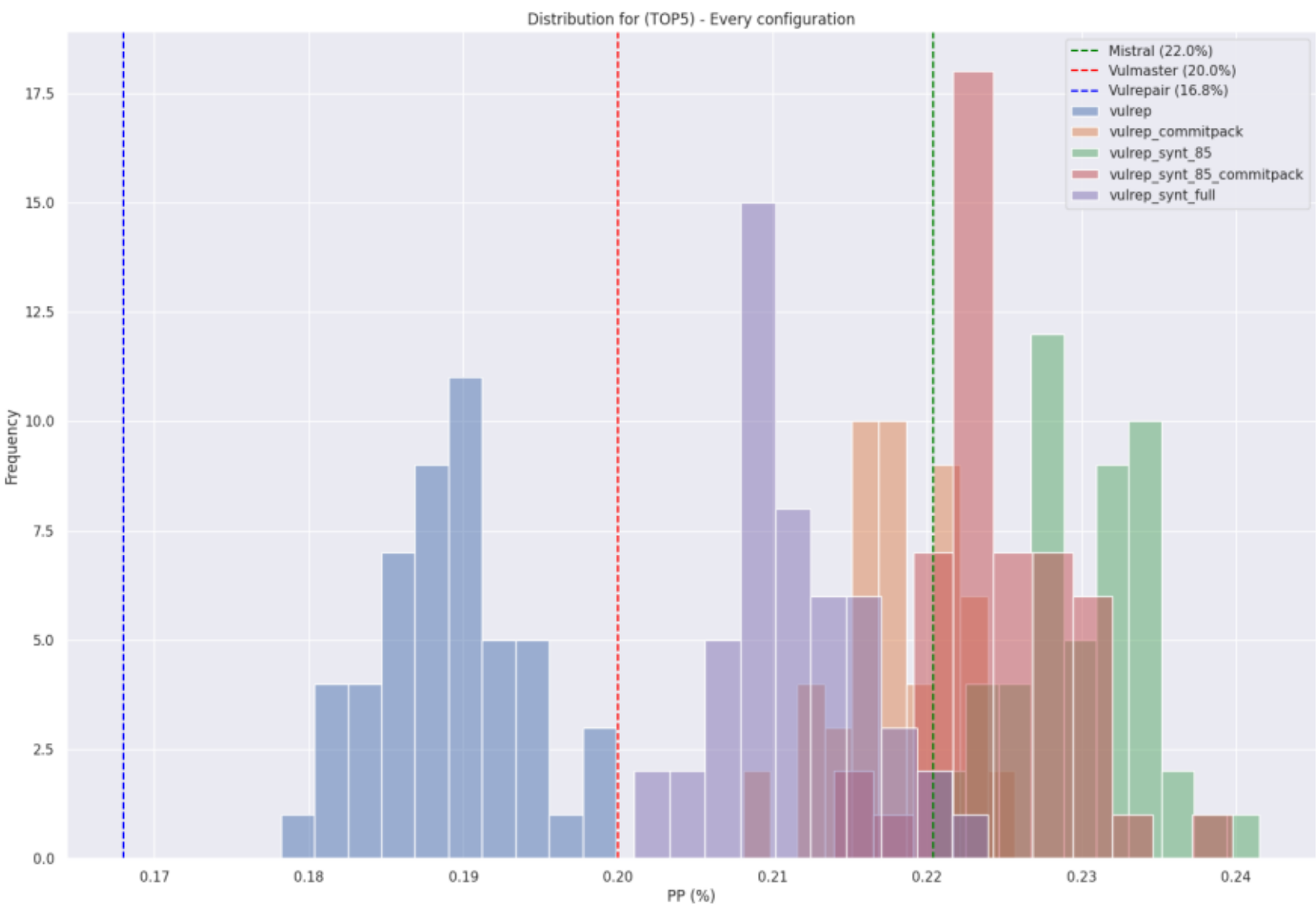

*Figure 2: Probability density distributions for Top@5 configurations*

## 5.3 Ablation study

To address RQ3, we investigate the sensitivity of our approach to its key design choices by examining two aspects: (1) the quality threshold used for filtering synthetic samples, and (2) the weights assigned to different evaluation criteria.

Due to computational constraints, these experiments used reduced configurations (single epoch training, 500-sample evaluation sets) with 25 repetitions per configuration to ensure statistical validity. All experiments were evaluated using the Top@5 setting.

The baseline configuration used the same reduced training setup but without any synthetic data, relying just on the VulRepair training set. While absolute performance values are lower due to the reduced experimental setup, relative differences between configurations remain informative.

Figure 3 presents the results of both ablation studies. For threshold analysis (left panel), we tested values from 7.0 to 9.0, finding that moderate thresholds (7.5-8.5) yielded optimal performance, with 7.5 achieving the highest mean PP% (19.58%). The threshold of 8.5 used in our main experiments (19.08%) proved near optimal, validating our choice. Higher thresholds (9.0) showed significant performance degradation (17.30%), likely due to excessive filtering, reducing dataset diversity. Importantly, all synthetic data configurations consistently outperformed the baseline without synthetic data (11.60%), demonstrating that incorporating synthetic data consistently improves APR performance regardless of the specific filtering parameters used.

For weight analysis (right panel), keeping the threshold fixed at 8.5, we compared our original weighting scheme against alternatives prioritizing individual criteria or using equal weights. We tested five different weight configurations: (1) *balanced* (our original), (2) *quality-focused,* (3) *correctness-focused,* (4) *security-focused,* and (5) *equal weights*. The original configuration achieved the best performance (19.08%), followed closely by quality-focused weights (18.78%). All weighted configurations significantly outperformed the baseline without synthetic data, confirming the effectiveness of quality filtering regardless of specific weight choices.

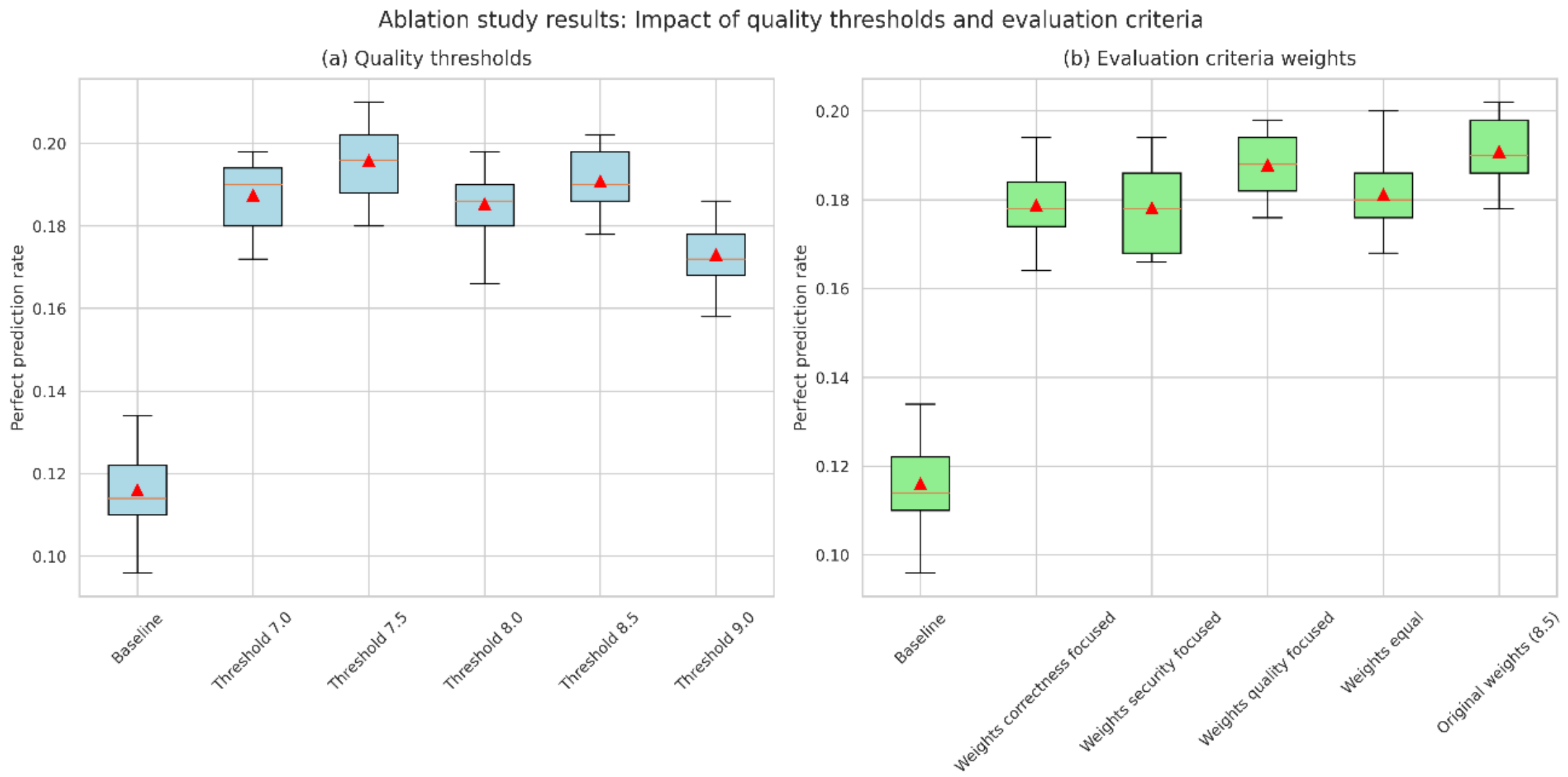

*Figure 3: Ablation study results showing PP% (Top@5) for different (a) quality thresholds and (b) evaluation criteria weights. Boxes show quartiles with means marked by triangles. The baseline shows performance using only VulRepair training data with reduced training configuration*

These ablation results answer RQ3: the approach is robust to variations in both filtering parameters. Threshold analysis shows that moderate values (7.5-8.5) yield near-optimal performance, with our chosen threshold of 8.5 performing within 0.5 percentage points of the best value (7.5). Only extreme thresholds (9.0) cause notable degradation due to excessive filtering that reduces dataset diversity. Weight analysis confirms that the original balanced weighting achieves the best results, but all alternative weight configurations still substantially outperform the baseline. Taken together, these findings indicate that the methodology is not sensitive to specific parameter choices within reasonable ranges, and that the benefits of synthetic data generation are consistent across different filtering configurations.

# 6 Discussion

Our experimental results provide empirical evidence for the effectiveness of synthetic data generation in enhancing automated program repair capabilities. This section examines the implications of these findings, contextualizes them within existing literature and acknowledge the limitations of our approach.

## 6.1 Implications and contributions

The experimental results demonstrate that quality-filtered synthetic data significantly enhances automated program repair capabilities. The filtered synthetic dataset (*vulrep_synt_85*) achieved approximately 2 percentage points higher performance than the unfiltered version (*vulrep_synt_full*) in both evaluation settings (Top@1 and Top@5), empirically validating that data quality outweighs quantity for training effective APR systems. This finding aligns with emerging principles in deep learning that emphasize training data quality over quantity (DeepSeek-AI, Liu, et al., 2024; Dubey et al., 2024).

The comparable performance between *vulrep_synt_85* and *vulrep_synt_85_commitpack* in Top@1 evaluations suggests that high-quality synthetic data may encapsulate much of the same repair knowledge present in real-world commit data. The statistically significant difference in Top@5 suggests that synthetic data could lead to more diverse patch generation, which is a valuable capability for practical applications where developers must choose among alternatives.

Our statistical validation through ANOVA and post-hoc Tukey's HSD tests establishes robust evidence for these improvements. Notably, the mean difference of 0.0549 between *vulrep_synt_85* and *baseline* in Top@1 represents a 47% relative improvement (17.18% vs 11.68%), surpassing gains from adding real commit data alone (0.0446). This rigorous statistical framework, uncommon in APR literature, sets a new methodological standard for empirical validation in the field.

Central to our methodology, the cross-model evaluation approach represents a significant innovation. Using multiple LLMs as both generators and evaluators, we established a consensus-based quality assessment that effectively mitigates individual model biases. This self-bootstrapping paradigm, where models generate and subsequently evaluate their own training data, demonstrates considerable potential for autonomous data curation in software engineering tasks. Additionally, this approach addresses the challenge of objective quality assessment in domains where ground truth is expensive to obtain.

When compared to existing systems, our experiments show that models trained with quality-filtered synthetic data achieve performance comparable to or exceeding previously reported benchmarks. Specifically, our approach achieves a 15% relative improvement over VulMaster (23.0% vs 20.0%) (Zhou, Kim, et al., 2024) and a 37% relative improvement over VulRepair (23.0% vs 16.8%) (Fu et al., 2022), even though our approach uses only 5 candidates versus their beam search with 50. This efficiency gain importantly suggests that improved training data quality can reduce reliance on exhaustive inference strategies, thus offering practical advantages for deployment.

To contextualize our work within the broader literature, our findings align with recent research on synthetic data generation addressing data scarcity (Liu et al., 2024; Nadas et al., 2025). In particular, the effectiveness of combining filtered synthetic data with real-world baselines supports established strategies for mitigating distribution shift (Li et al., 2024). Additionally, our multi-criteria evaluation framework extends beyond execution-based validation common in code generation (Chen et al., 2024; Huang et al., 2022), thereby providing semantic quality assessment valuable when test suites are unavailable. This holistic evaluation approach represents a significant methodological contribution.

While our evaluation is limited to vulnerability repair on the VulRepair benchmark, the underlying methodology may be applicable to other software engineering tasks facing similar data limitations, such as code translation, refactoring, and documentation generation, where high-quality paired examples are scarce. Validating this hypothesis across different tasks and benchmarks remains an important direction for future work.

## 6.2 Comparison with code-specialized LLMs

To investigate whether code-specialized LLMs produce higher-quality synthetic data than general-purpose models, we conducted a comparative study. We generated samples using five Code LLMs: CodeLlama-70b-Instruct (Rozière et al., 2024), DeepSeek-Coder-V2-Lite (DeepSeek-AI, Zhu, et al., 2024), Qwen2.5-Coder-32B (Hui et al., 2024), Qwen3-Coder-30B (Team, 2025), and Devstral-Small[2], requesting 5,000 samples from each model. Valid outputs from all five models were merged into a single code LLM dataset. These samples were first evaluated by our six general-purpose evaluators, and additionally by the same five code LLMs acting as evaluators to control for potential bias favoring general-purpose generators. After quality filtering (threshold ≥8.5), we applied stratified sampling by language, bug type, and score distribution to obtain three comparable 5,000 sample datasets: one from general-purpose generators (sampled from the same filtered dataset used in our main experiments), and two from code LLM generators (filtered using either general-purpose or code LLM evaluators). Due to computational constraints, downstream evaluation employed a reduced setup similar to our ablation study: single epoch training, 500 sample evaluation sets, and 25 repetitions per configuration.

A notable finding during generation was the substantial variation in format compliance among Code LLMs (Table 17). While some models achieved near-perfect validity rates (Devstral: 99.96%, DeepSeek-Coder-V2: 99.94%), others showed significant compliance issues (CodeLlama-70b: 27.58%, Qwen2.5-Coder-32B: 46.18%). These failures primarily occurred because these models refused to generate security-related vulnerabilities, responding with refusal messages instead of the expected code.

| Model | Valid Samples | Validity (%) |
|---|---|---|
| Devstral-Small-2507 | 4,998 | 99.96% |
| Qwen3-Coder-30B-A3B-Instruct | 4,955 | 99.10% |

[2] https://mistral.ai/news/devstral

| | | |
|---|---|---|
| DeepSeeek-Coder-V2-Lite-Instruct | 4,997 | 99.94% |
| Qwen2.5-Coder-32B-Instruct | 2,309 | 46.18% |
| CodeLlama-70b-Instruct-hf | 1,379 | 27.58% |

*Table 17: Code LLM format validity rates for synthetic data generation*

Cross-model evaluation using our six general-purpose evaluators showed that samples from general-purpose generators achieved higher weighted quality scores than code LLM samples (8.58 vs 8.37, $p<0.001$), though with considerable variance among code LLMs (scores ranging 7.50–8.95). When code LLM samples were evaluated by code LLM evaluators, scores were slightly higher (8.54), suggesting modest self-preference bias. Notably, Qwen3-Coder-30B, the most recently released model, achieved the highest individual score (8.95-9.06, depending on evaluator type), suggesting that newer code-specialized models may be addressing the instruction-following limitations observed in earlier models.

Downstream evaluation on the VulRepair test set (Table 18) confirmed that training on general-purpose synthetic data yielded better results. Pairwise Mann-Whitney U tests revealed that general-purpose data significantly outperformed code LLM data in Top@1 (11.18% vs 10.51%, $p=0.045$), while Top@5 showed no significant difference (16.48% vs 15.87%, $p=0.140$). We also evaluated whether using code LLM evaluators to filter code LLM-generated data could reveal potential bias when evaluating with the same models that generated the dataset, but this approach showed no significant improvement in downstream evaluation over using general-purpose evaluators (10.22% vs 10.51%, $p=0.559$). Importantly, all synthetic datasets substantially outperformed the baseline trained only on VulRepair data.

| Dataset | Top@1 | Top@5 |
|---|---|---|
| Baseline | 7.06% | 11.60% |
| General-purpose synthetic | **11.18%** | **16.48%** |
| Code LLM synthetic (Gen eval) | 10.51% | 15.87% |
| Code LLM synthetic (Code eval) | 10.22% | 16.07% |

*Table 18: Downstream evaluation (PP%) comparing synthetic data sources. "Code eval" indicates samples filtered using code LLM evaluators, while "Gen Eval" indicates samples filtered using generalist models. Best results in bold.*

These results suggest that for APR tasks, general-purpose models with strong code capabilities are better suited for synthetic data generation than code-specialized models. This advantage may be attributed to better instruction-following capabilities, including willingness to generate security-related bug patterns and more consistent adherence to structured output formats.

## 6.3 Threats to validity

Several factors may limit the validity of our findings, beginning with internal validity concerns. The correctness of LLM-based evaluation represents a primary consideration, as while cross-model consensus mitigates individual biases, all evaluator models share similar training paradigms and could consequently exhibit correlated errors. Furthermore, our weighted scoring criteria (correctness 0.3, code quality 0.2, etc.) were determined through expert judgment rather than empirical optimization, potentially affecting sample selection. Additionally, the 8.5 quality threshold, though demonstrably effective, was chosen pragmatically rather than through systematic exploration.

Temperature settings during generation (0.7) and evaluation (0.2) could also influence results, though internal testing showed minimal sensitivity to these parameters.

The scope of our evaluation also introduces certain limitations. Our experiments exclusively used the VulRepair test set, focusing on C/C++ vulnerabilities, despite generating multi-language synthetic data. Consequently, generalization to other benchmarks, programming languages, or bug types remains unverified. Moreover, the selected LLMs (27B-72B parameters) represent current frontier models, and results may not transfer to smaller models or future architectures. This is particularly relevant as our methodology assumes access to multiple high-quality LLMs for cross-evaluation, which may limit applicability in resource-constrained settings. While the specific bug categories and programming languages chosen were comprehensive, they may not fully represent all real-world repair scenarios. These factors collectively suggest that additional validation across diverse contexts would strengthen claims of generalizability.

Furthermore, while our preliminary human validation study on 100 stratified samples demonstrates alignment between LLM consensus and human judgment, the study used a single expert evaluator and represents only a small fraction of our dataset (20,000+ samples across 12 languages). More extensive validation with multiple evaluators and inter-rater reliability analysis would strengthen confidence in the generalizability of these findings across the full 20,000+ sample dataset spanning 12 programming languages. Additionally, our filtering methodology does not employ automated syntactic validators (compilers, linters) or execution-based testing as quality criteria. While post-hoc validation on a sample demonstrated high syntactic correctness (97.7% across five languages), comprehensive validation across all 12 languages and execution-based testing remain unexplored.

Finally, the evaluation metrics themselves present additional considerations. PP% measures exact match with ground truth, which may not adequately reflect real-world repair success where functionally equivalent alternatives exist. This metric could undervalue models generating correct but syntactically different fixes. The assumption that higher weighted scores correlate with better repair quality similarly lacks empirical validation against developer preferences or deployment success. This becomes particularly significant when considering that our focus on vulnerabilities rather than general bugs may limit insights about broader applicability, as security fixes often follow more structured patterns than general debugging. Taken together, these limitations suggest that while our metrics provide valuable insights, they may not fully capture the complexity of real-world program repair.

# 7 Conclusion

This research demonstrates the effectiveness of synthetic data generation for automated program repair, addressing a critical gap in current APR methodologies. The generative capabilities of LLMs were leveraged to create diverse, high-quality training examples across multiple programming languages and bug types, resulting in remarkable improvements in repair performance. The results empirically validate that quality-filtered synthetic data can boost repair accuracy beyond what is achievable with a manually gathered dataset, with a filtered dataset (threshold of 8.5) yielding statistically significant improvements over both the baseline and unfiltered synthetic datasets.

The cross-model evaluation methodology introduced in this work represents a novel approach to quality assessment in synthetic data generation. Statistical analyses confirmed the significance of the improvements achieved through this approach, with ANOVA and post-hoc Tukey's HSD tests revealing statistically significant differences between training configurations.

The comparative analysis with existing APR systems further validates the effectiveness of our system. The best-performing configurations achieved superior results compared to established benchmarks like VulMaster (Zhou, Kim, et al., 2024) and VulRepair (Fu et al., 2022), despite using a less computationally intensive sampling approach rather than beam search. This difference in performance underscores the potential of high-quality synthetic data to fundamentally improve model capabilities, reducing reliance on exhaustive search strategies during inference. The consistent performance across multiple experimental runs, confirmed through probability density visualizations, proves the reliability and stability of the proposed approach in real-world applications.

The promising results achieved with moderate-sized language models (7B parameters) suggest potential for scaling to larger models, which may capture even more nuanced repair patterns and further improve performance. This scaling can affect not only the fine-tuned models, but also those used for synthetic data generation, as larger models may capture more subtle repair patterns and contribute to further improving training data quality.

Regarding model selection, general-purpose LLMs outperformed code-specialized models for synthetic data generation in APR tasks, especially those involving security vulnerabilities. This result appears attributable to differences in instruction-following behavior, as code-specialized models more frequently refused to generate security-related bug patterns.

Our preliminary human validation study on 100 samples suggests that LLM-based evaluation aligns with human judgment, but larger-scale validation with multiple expert evaluators and inter-rater reliability analysis would strengthen these findings. Additionally, deeper analysis of inter-model agreement patterns and score variance would provide further insights into potential evaluator biases and inform refinements to the scoring criteria and weighting schemes.

Several other promising research directions emerge from this work. Exploration of more sophisticated filtering techniques, potentially incorporating non-LLM oracles such as compilers, static analyzers, and execution-based testing on generated samples, or formal verification methods, could further improve data quality and break potential circularity in the evaluation pipeline. Investigation of domain-specific fine-tuning for particularly challenging bug types, like concurrency issues, may yield improvements in areas where current approaches demonstrate reduced effectiveness. Development of hybrid approaches that combine synthetic data generation with targeted mining of real-world repositories could leverage the strengths of both worlds. Evaluation on general-purpose bug repair benchmarks such as Defects4J would demonstrate the generalizability of our approach beyond security vulnerabilities. Larger-scale human validation studies with multiple expert evaluators and inter-rater reliability analysis would further strengthen confidence in the LLM-based quality assessment methodology. Additionally, extension of this framework to other programming languages and bug types not covered in the current study would enhance generalizability. Another promising direction involves providing base code samples to LLMs during synthetic generation, potentially improving the realism of generated bug scenarios while maintaining diversity through careful sampling of the base code contexts.

**Acknowledgements**

This work was supported by the project “Tecnologías Inteligentes para la Fabricación, el diseño y las Operaciones en entornos iNdustriales” (TIFON, PLEC2023-010251) through the call Proyectos de I+D+i en líneas estratégicas - Transmisiones 2023 and by the project "Reparación automática de código fuente mediante modelos generativos de Procesamiento de Lenguaje Natural" (RACO-NLP, SBPLY/23/180225/000063) through the call Ayudas para la realización de proyectos de investigación científica y transferencia de tecnología de Castilla-La Mancha 2023.

**Author contributions: CRediT**

**David De-Fitero-Dominguez**: Conceptualization, Data curation, Methodology, Software, Investigation, Writing – Original Draft. **Antonio Garcia-Cabot**: Project administration, Funding acquisition, Supervision, Investigation, Writing – Original Draft, Resources. **Eva Garcia-Lopez**: Project administration, Formal analysis, Validation, Funding acquisition, Supervision, Investigation, Writing – Original Draft.

# A Appendix

## A.1 Prompts

### A.1.1 Data generation prompt

The prompt template used for generating synthetic bug/fix pairs is structured as follows:

You are an expert software developer and secure coding specialist. Your task is to generate pairs of code snippets that demonstrate common programming bugs and their corresponding fixes. Each pair should include:

- **Error Description:** A brief explanation of the specific bug or vulnerability in the code.
- **Buggy Code:** A detailed code snippet that contains the bug. Ensure the code realistically reflects common development mistakes or vulnerabilities.
- **Fixed Code:** The corrected version of the buggy code with the identified issue resolved.

## Requirements:

- **Bug or Vulnerability Type:** The buggy code should contain a realistic programming bug or vulnerability of the following type: {bug}.
- **Programming Languages:** The code should be written in {language}.
- **Realism and Complexity:** The code should be at least 50 lines long and resemble real-world scenarios, including realistic functions, classes, or modules that might appear in software projects.
- **Output Format:** Ensure your response strictly follows this format:

<error_description>

Brief description of the bug or vulnerability

</error_description>

<buggy_code>

Code snippet containing the bug

</buggy_code>

<fixed_code>

Corrected code snippet

</fixed_code>

{one-shot Java example demonstrating a race condition fix, omitted for brevity}

Do not include backticks to represent the source code. Do not generate any additional text or information. Only the specified output format is accepted. A single pair of code snippets is required.

### A.1.2 Patch quality evaluation prompt

The evaluation system employs a multi-dimensional scoring approach using the following prompt:

You are an expert software developer and secure coding specialist. Your task is to evaluate pairs of code snippets that demonstrate common programming bugs and their corresponding fixes. Each pair includes:

- **Error Description:** A brief explanation of the specific bug or vulnerability in the code.
- **Buggy Code:** A detailed code snippet that contains the bug.
- **Patch:** The generated patch that fixes the identified issue in the buggy code.

Your evaluation should include:

- **Correctness:** Whether the fixed code correctly resolves the issue described in the error description.
- **Code Quality:** Whether the fixed code maintains or improves the overall code quality, including readability, maintainability, and adherence to best practices.
- **Security:** Whether the fixed code addresses any security vulnerabilities present in the buggy code.
- **Performance:** Whether the fixed code maintains or improves the performance of the original code.
- **Completeness:** Whether the fixed code fully addresses all aspects of the bug or vulnerability described.

## Requirements:

- **Rating:** Assign a score from 0 to 10 for each criterion.
- **Output Format:** Ensure your response strictly follows this JSON format:

```
{{
  "correctness": x,
  "code_quality": x,
  "security": x,
  "performance": x,
  "completeness": x
```

}}

Provide only a JSON object with ratings, no additional text or information.

This is the sample you need to evaluate:

<error_description>

{error_description}

</error_description>

<buggy_code>

{buggy_code}

</buggy_code>

<patch>

{patch}

</patch>

## A.2 Generated examples

### A.2.1 High-quality example (overall score: 9.03/10)

**Sample ID**: a31f70e7f98575eebe0b43f06dcdcf8f

**Bug type**: Off-by-one errors

**Language**: Ruby

**Evaluation scores**:

- Correctness: 9.83
- Code quality: 9.50
- Security: 10.00
- Performance: 10.00
- Completeness: 9.83

**Error description**: Off-by-one error in loop indexing, causing the last element to be skipped.

Key change (within a 16-line Ruby class):

**Buggy:** inclusive range causes out-of-bounds access

```
for i in 0..@array.length
```

**Fixed:** exclusive range correctly iterates array indices

```
for i in 0...@array.length
```

### A.2.2 Lower-quality example (overall score: 6.88/10)

**Sample ID**: 704f04792087991c9269de65b89d9c1a

**Bug type**: Improper error handling

**Language**: Kotlin

**Evaluation scores**:

- Correctness: 7.17
- Code quality: 6.17
- Security: 6.33
- Performance: 8.83
- Completeness: 7.00

**Error description**: Improper error handling in a file reading operation that leads to uncaught exceptions.

Key change (within a 17-line Kotlin program):

**Buggy:** catches only FileNotFoundException, missing other potential exceptions

```
} catch (e: FileNotFoundException) {
    println("File not found: ${e.message}")
```

**Fixed:** catches general Exception for broader error handling

```
} catch (e: Exception) {
    println("An error occurred: ${e.message}")
```